\documentclass[11pt,twoside]{article}
\usepackage{asp2010}

\resetcounters

\begin{document}

\title{RAMSES-CH: A New Chemodynamics Code}
\author{C.~Gareth~Few,$^1$ St\'{e}phanie~Courty,$^2$ and Brad~K.~Gibson,$^1$
\affil{$^1$Jeremiah Horrocks Institute, University of Central Lancashire, Preston, PR1~2HE, UK}
\affil{$^2$Centre de Recherche Astrophysique de Lyon, Ecole Normale Sup\'{e}rieure de Lyon, Lyon, F-69007, France}}

\begin{abstract}
We present a new chemodynamical code based on the adaptive mesh refinement code \textsc{ramses}. The new code uses Eulerian hydrodynamics and N-body dynamics in 
a cosmological framework to trace the production and advection of several chemical species. It is the first such code to follow the self-consistent evolution of
chemical elements in cosmological volumes while maintaining sub-kiloparsec resolution. The code will be used to simulate disk galaxies and explore the influence 
of chemical evolution models and star formation on galactic abundance ratios.
\end{abstract}

There are numerous cosmological codes on the market but studies of chemical evolution (CE) are limited 
to smoothed particle hydrodynamics with a dearth of Eulerian implementations that include a detailed 
chemical evolution model. With the aim of providing 
a complementary approach to existing CE codes we present a fully cosmological, CE code with an adaptive 
mesh refinement hydrodynamics scheme that traces the formation and subsequent 
evolution of H, He, C, N, O, Ne, Mg, Si and Fe. The model presented here uses a \citet{kroupa93} initial mass 
fraction (IMF) and a SNIa delayed time distribution inspired by \citet{kawata03}. 
Stellar lifetimes are taken from \citet{kodama97}, SNIa yields from \citet{iwamoto99}, SNII yields from 
\citet{ww95} and AGB stellar wind yields from \citet{vdhoek97}.

\subsection*{Results}
Our code is applied to cosmological simulations of disk galaxies achieving a resolution of 
436 pc. The galaxy presented is a field spiral galaxy of total mass 6.8$\times$10$^{11}$M$_\odot$. This galaxy 
will be used as a fiducial model for a series of test runs using different IMFs and SNIa models to ascertain the influence 
that each of these ingredients has on the abundance ratios, metallicity gradients, morphology and kinematics of the galaxy.
\articlefiguretwo{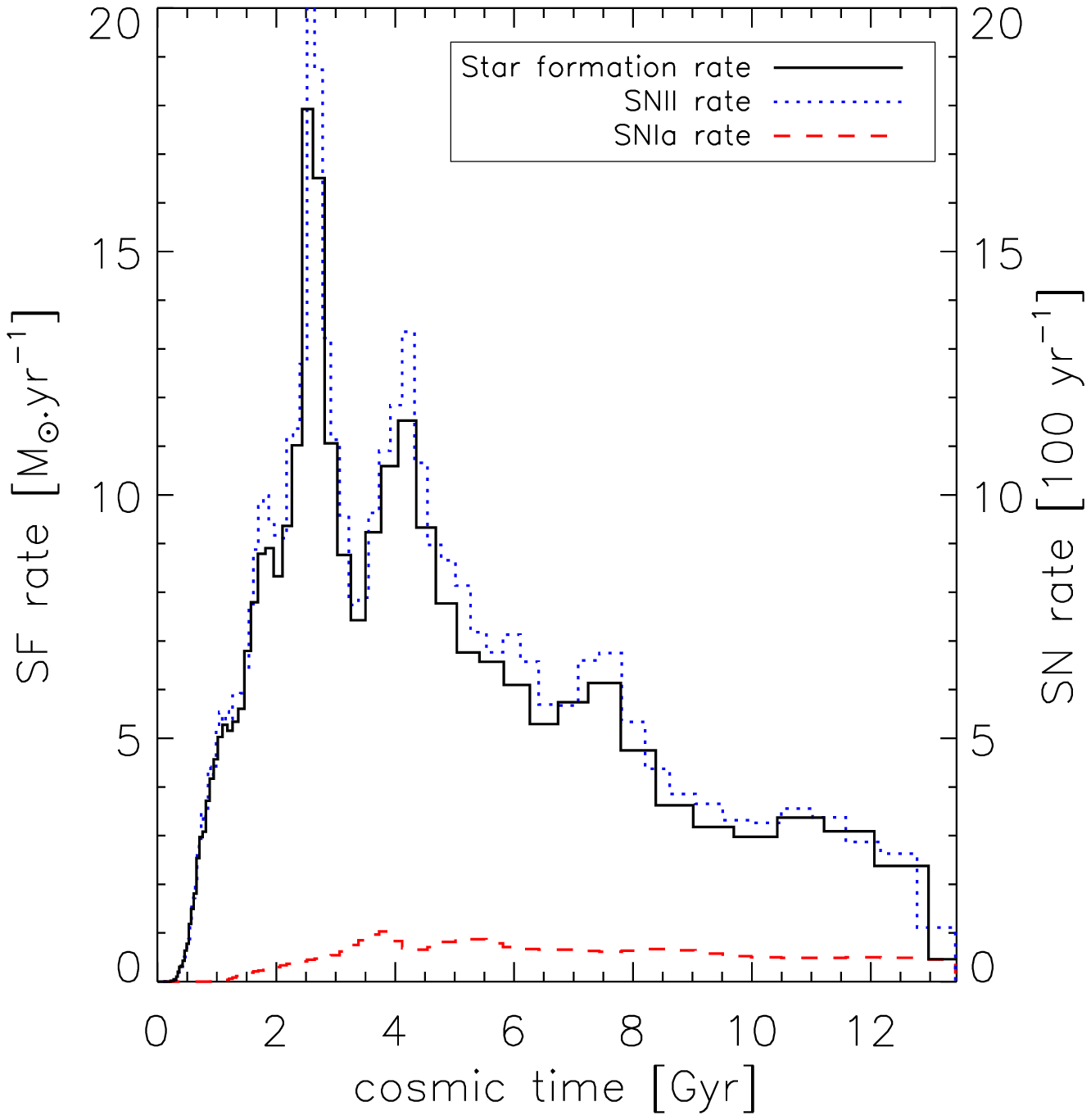}{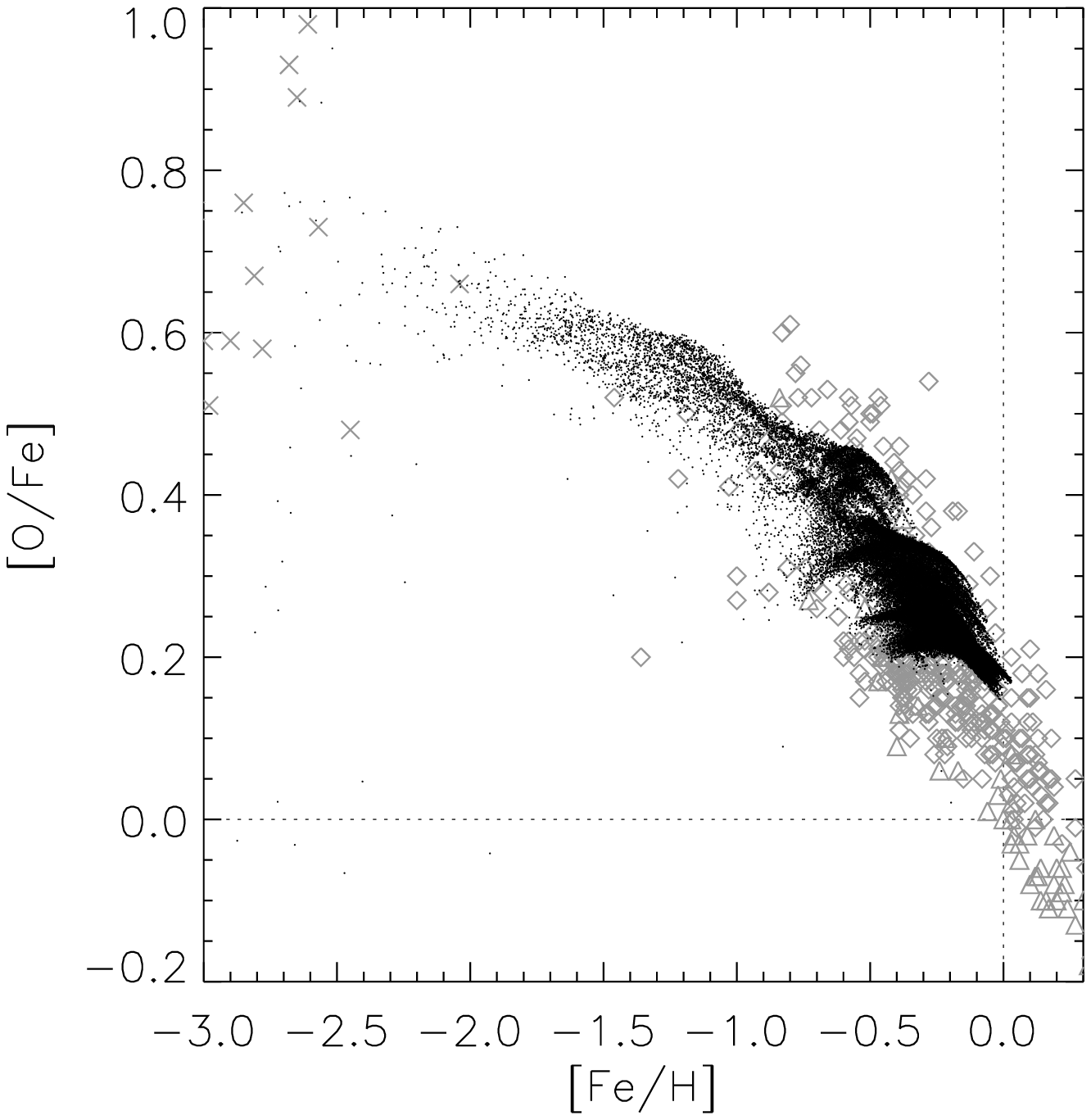}{snandabundance}{Star formation and SN rates (left panel). The z=0 SN rates are SNII=0.763$SNuM$ and 
SNIa=0.096$SNuM$ (SN per century per $10^{10}M_\odot$) and compare favourably with those observed by \citet{mannucci08} of 
0.65$^{+0.16}_{-0.13}SNuM$ (SNII) and $0.14^{+0.5}_{-0.04}SNuM$ for (SNIa) in field Sbc/d galaxies. 
An example of [O/Fe] vs. [Fe/H] (right panel) for the particular CE model used in this run. Observational data is plotted in grey, 
\citet{cayrel04} (crosses), F/G/K dwarfs from \citet{ramirez07} (diamonds) and solar neighborhood F/G stars from \citet{bensby04} (triangles).
}
A degree of success is achieved in fitting observations of the Milky Way disk (however we stress that this galaxy is by no means a 
Milky Way clone) but most parameter combinations are too $\alpha$-rich at the high metallicity end of the 
distribution. It is believed that this can be traced to the relatively low SNIa rate and future runs will explore this in more detail.

\subsection*{Summary}

We present the first of what will become a suite of Eulerian cosmological disk galaxy simulations with CE 
and sub-kpc resolution. The first simulations show a good agreement of the SN rates with observations and an improved 
rotation curve with respect to their counterparts created with the standard feedback mechanism. A great deal of variation 
in abundance ratios is seen under changes in initial mass function slope, upper mass limit and SNIa delayed 
time distribution. Future work will explore a full range in parameter space to constrain the CE of disk galaxies.

\acknowledgements We thank Romain Teyssier, Daisuke Kawata and Francesco Calura for significant contributions to this work.

\bibliography{aspauthor}
\end{document}